\documentclass[conference]{IEEEtran}
 \IEEEoverridecommandlockouts

\usepackage{ifpdf}
\usepackage{cite}
\usepackage{color}
\ifCLASSINFOpdf
 \usepackage[pdftex]{graphicx}
\else
   \usepackage[dvips]{graphicx}
   \DeclareGraphicsExtensions{.eps}
\fi
\usepackage[cmex10]{amsmath}
\usepackage{algorithmic}
\usepackage{array}
\usepackage{mdwmath}
\usepackage{mdwtab}
\usepackage{eqparbox}
\usepackage{fixltx2e}
\usepackage{stfloats}
\usepackage{url}
\usepackage{epstopdf}
\usepackage{enumerate}
\usepackage{amsmath, amsthm, amssymb}

\newtheorem{definition}{Definition}
\newtheorem{theorem}{Theorem}

\newtheorem{claim}{Claim}
\newtheorem{remark}{Remark}
\newtheorem{example}{Example}

\begin{document}

\title{Information Theoretic Limits of Data Shuffling for Distributed Learning\vspace{-28pt}}

\vspace{-20pt}
\author{Mohamed Adel Attia \hspace{25pt} Ravi Tandon\\
  Department of Electrical and Computer Engineering\\
 University of Arizona, Tucson, AZ, 85721\\
Email: \emph{\{madel, tandonr\}}@email.arizona.edu}

\maketitle

\begin{abstract}
Data shuffling is one of the fundamental building blocks for distributed learning algorithms, that increases the statistical gain for each step of the learning process.
  In each iteration, different shuffled data points are assigned by a central node to a distributed set of workers to perform local computations, which leads to communication bottlenecks. 
The focus of this paper is on formalizing and understanding the fundamental information-theoretic tradeoff between storage (per worker) and the worst-case communication overhead for the data shuffling problem. 
We completely characterize the information theoretic tradeoff for $K=2$, and $K=3$ workers, for any value of storage capacity, and show that increasing the storage across workers can reduce the communication overhead by leveraging coding. We propose a novel and systematic data delivery and storage update strategy for each data shuffle iteration, which preserves the structural properties of the storage across the workers, and aids in minimizing the communication overhead in subsequent data shuffling iterations.
\end{abstract}

\vspace{-5pt}

%

\section{Introduction}
\label{sec:Introduction}

Distributed computing systems for large data-sets have gained a lot of interest recently as they enable the processing of data-intensive tasks for machine learning, model tracing, and data analysis over a large number of commodity machines, and servers (e.g., Apache Spark \cite{ZaChFrShSt2010}, and MapReduce \cite{DeGh2004}). Generally speaking, a master node, which has the entire data-set, sends data blocks to be processed at distributed worker nodes.  The workers subsequently respond with locally computed functions to the master node for the desired data analysis. This enables the processing of many terabytes of data over thousands of distributed servers in a time efficient manner.


One of the core elements in distributed learning algorithms is \textit{data shuffling} \cite{S-arxiv-2016}.  Before each iteration of the learning process, the entire data is randomly shuffled before being assigned to the worker nodes. This shuffling operation enables the worker nodes to process different data batches at each iteration, which presents large statistical gains \cite{GuOzPa:2015,IoSz:2015}. The statistical advantages provided by data shuffling come at the unavoidable cost of the communication overhead between the master and the worker nodes which must be incurred for every shuffling iteration. Thus, there exists a fundamental tradeoff between the communication overhead and the storage capacity of each worker node. To exemplify this tradeoff, consider two extreme scenarios: an ideal scenario in which  the storage at the distributed workers is large enough to store the entire data-set, thus no communication has to be done from the master node for any shuffle. On the other extreme, when the storage is just enough to store the batch under processing, the communication load is expected to be maximum.

The focus of this paper is on formalizing and understanding this fundamental information-theoretic tradeoff between storage and the worst-case communication overhead for the data shuffling problem. Each iteration of data shuffling can be divided into two phases: \textit{data delivery}, and \textit{storage update}. In the data delivery phase, depending on the shuffled data points, the master node communicates a function of the data to the workers, so that  each worker obtains its assigned data points. 
The second phase is termed as the storage update phase, which as shown in this paper is extremely critical in reducing the communication overhead of subsequently shuffling iterations. 
We next summarize the main contributions of this paper:


\noindent \hspace{4pt}$\bullet$ We first present an information-theoretic formulation of the data shuffling problem involving both data delivery and update phases, accounting for the respective constraints and formalizing the tradeoff between worst-case communication overhead and the storage capacity of distributed workers. 
\vspace{-1pt}
%

\noindent \hspace{4pt} $\bullet$ We also completely characterize this tradeoff for $K=2$ and $K=3$ workers, for any value of storage capacity. One of the most interesting aspects of the result for $K=3$ worker problem is the design of data delivery and storage update algorithms. In particular, for data delivery phase, we show that transmitting coded data from the master node to the workers can significantly reduce the communication overhead.  More interestingly, the proposed storage update algorithm maintains the structural properties of the storage at the workers over time. This structural invariance placement is extremely critical in leveraging the gains of coding for different shuffles.   

\textbf{Related Work:} 
In the past few years, there has been a flurry of research acitivity in understanding the benefits of coding for caching starting from the work of Maddah-Ali and Niesen \cite{MaNi:2014} who showed that exploiting multi-casting opportunities by coding can reduce the communication for caching. In \cite{LiMaAv:2015}, coding for MapReduce was proposed in order to reduce the communication cost between mappers and reducers, however the underlying focus of  \cite{LiMaAv:2015} is significantly different than the problem considered in this work, where we care about the communication between the master node and the workers. The paper most closely related to this work is \cite{LeLaPePaRa:2015}, where the idea of coding for data shuffling problem is presented to reduce the communication overhead between the master node and worker nodes.  \cite{LeLaPePaRa:2015} provides a probabilistic scheme of leveraging coding based on a random storage placement.
In contrast to \cite{LeLaPePaRa:2015}, in this paper we provide a deterministic and systematic storage update scheme, which increases the coding opportunities in the delivery phase. The underlying metric used here is the worst-case communication cost over all the possible shuffles, unlike the average cost considered in \cite{LeLaPePaRa:2015}. Finally, we also present the first information theoretic lower bounds on the communication overhead for the data shuffling problem. 


\section{System Model}
\label{sec:System}


We assume a master node which has access to the entire data-set $A=[x_1^T,x_2^T,\ldots,x_N^T]^T$ of size $Nd$ bits, i.e., $A$ is a matrix containing $N$ data points, denoted by $x_1,x_2,\ldots,x_N$, where $d$ is the dimensionality of each data point. Treating $A$, and its data points $x_n$ as random variables, we therefore have the entropies of these random variables as
\begin{align}
\label{eq:data-set}
&H(A)=N\times H(x_n)=Nd, \quad \forall n\in \{1, 2, \ldots, N\}.
\end{align}

At each iteration, indexed by $t$, the master node divides the data-set $A$ into $K$ data batches given as $A^{t}_{1}, A^{t}_{2}, \ldots, A^t_K$, where the batch $A^t_k$ is designated to be processed by worker $w_k$, and these batches correspond to the random permutation of the data-set, $\pi^t:A\rightarrow\{A^{t}_{1}, \ldots, A^t_K\}$.  Note that these data chunks are disjoint, and span the whole data-set, i.e.,
\begin{subequations}
\label{eq:data-batches}
\begin{align}
&A^t_i \cap A^t_j = \phi, \quad \forall i\neq j,\\
&A^t_1 \cup A^t_2 \cup \ldots \cup A^t_K =A, \quad \forall t.\label{eq:data-partitions}
\end{align}
\end{subequations}
Hence, the entropy of any batch $A^t_k$ is given as
\begin{align}
\label{eq:data-batches2}
H(A^t_k)= \frac{1}{K} H(A)= \frac{N}{K}d,\quad \forall k\in\{1,\ldots,K\}.
\end{align}

After getting the data batch, each worker locally computes a function (as an example, this function could correspond to the gradient or sub-gradients of the data points assigned to the worker). The local functions from the $K$ workers are processed subsequently at the master node. We assume that each worker $w_k$ has a storage $Z^t_k$ of size $Sd$ bits, for a real number $S$. 
For processing purposes, the assigned data blocks are needed to be stored by the workers, therefore, each worker $w_k$ must at least store the data block $A^t_k$ at time $t$. If we consider $Z^t_k$ as a random variable then the storage constraint is given by
\begin{align}
\label{eq:cache-storage}
H(Z^t_k)=Sd \geq H(A^t_k),\qquad \forall k\in\{1,\ldots,K\}.
\end{align}

According to (\ref{eq:data-batches2}) and (\ref{eq:cache-storage}), we get the \textit{minimum storage per worker} $S \geq \frac{N}{K}$.
We also have the \textit{processing constraint} as
\begin{equation}
\label{eq:cache-min-desired}
H(A^{t}_k|Z^{t}_k)=0,\quad \forall k\in\{1,\ldots,K\}.
\end{equation}

In the next epoch $t+1$, the data-set is randomly reshuffled at the master node according to a random permutation $\pi^{t+1}: A\rightarrow\{A^{t+1}_1, A^{t+1}_2, \ldots, A^{t+1}_K\}$. The main communication bottleneck occurs during  {{\textit{Data Delivery}}} since the master node needs to communicate the new data batches to the workers.
Trivially, if the storage (per worker) exceeds $Nd$ bits, i.e., $S\geq N$, then each worker can store the whole data-set, and no communication has to be done between the master node and the workers for any shuffle. Therefore from the constraint on minimum storage per worker, we can write the possible range for  storage  as $\frac{N}{K}\leq S \leq N$.

We next proceed to describe the data delivery mechanism, and the associated encoding and decoding functions. 
The main process can be divided into 2 phases, namely the data delivery phase and the storage update phase as described next: 
\subsection{Data Delivery Phase}
At time $t+1$, the master node sends a function of the data batches for the subsequent shuffles $(\pi_t,\pi_{t+1})$, $X_{(\pi_t,\pi_{t+1})} = \phi(A^{t}_1, \ldots, A^{t}_K,A^{t+1}_1,\ldots, A^{t+1}_K)=\phi_{(\pi_t,\pi_{t+1})}(A)$ over the shared link, where $\phi$ is the data delivery encoding function
\begin{equation}
\phi: \left[2^{\frac{N}{K}d}\right]^{2K} \rightarrow [2^{R_{(\pi_t,\pi_{t+1})}d}],
\end{equation}
where $R_{(\pi_t,\pi_{t+1})}$ is the rate of the shared link based on the shuffles $(\pi_t,\pi_{t+1})$.
Therefore, we have 
 \begin{align}
\label{eq:transmit-load}
 H\left(X_{(\pi_t,\pi_{t+1})}|A\right)=0,\quad H\left(X_{(\pi_t,\pi_{t+1})}\right) =R_{(\pi_t,\pi_{t+1})}d.
 \end{align}
Each worker $w_k$ should decode the desired batch $A^{t+1}_k$ out of the transmitted function $X_{(\pi_t,\pi_{t+1})}$, and the data stored in the previous time slot denoted as $Z^{t}_k$. Therefore, the desired data is given by $A^{t+1}_k =\psi(X_{(\pi_t,\pi_{t+1})}, Z^{t}_k)$, where $\psi$ is the decoding function at the workers
\begin{equation}
\psi: [2^{R_{(\pi_t,\pi_{t+1})}d}]\times [2^{Sd}]\rightarrow [2^{\frac{N}{K}d}],
\end{equation} 
which can be written in terms of a \textit{decodability constraint}, at each worker as follows
\begin{equation}
\label{eq:decoding-const}
H\left(A^{t+1}_k|Z^{t}_k, X_{(\pi_t,\pi_{t+1})}\right)=0, \quad \forall k\in\{1,\ldots,K\}.
\end{equation}

\subsection{Storage Update Phase}
At each iteration, every worker updates its stored content as follows: the new storage content $Z^{t+1}_k$ is a function of the old storage content $Z^{t}_k$ as well as transmitted function $X_{(\pi_t,\pi_{t+1})}$, i.e., $Z^{t+1}_k=\mu(X_{(\pi_t,\pi_{t+1})},Z^{t}_k)$, where $\mu$ is the update function
\begin{equation}
\mu: [2^{R_{(\pi_t,\pi_{t+1})}d}]\times [2^{Sd}]\rightarrow [2^{Sd}].
\end{equation}
This implies the following \textit{storage-update constraint}
\begin{equation}
\label{eq:cache-update}
H(Z^{t+1}_k|Z^{t}_k, X_{(\pi_t,\pi_{t+1})})=0,\quad \forall k\in\{1,\ldots,K\}.
\end{equation}

The excess storage, if any, can be used to store opportunistically a function of the remaining data batches. Since the shuffling process at each time is done randomly, all the remaining batches are of equal importance. Consequently, the amount of excess storage, given by $(S-\frac{N}{K})d$ bits, is divided equally among the remaining $K-1$ batches. For the scope of this work, we assume that the placement of the excess storage is uncoded, which means that $\frac{(S-\frac{N}{K})}{K-1}d$ bits of the excess storage are dedicated to store a function of only one of the remaining $K-1$ batches.
We give the notation $A^{t+1}_{i,k}$, where $i\neq k$, as the part of data that worker $w_k$ stores about $A^{t+1}_i$ in the excess storage at  time $t+1$.
Considering $A^{t+1}_{i,k}$ as a random variable, then 
 \begin{align}
 \label{eq:excess-storage}
  H(A^{t+1}_{i,k}) =\left(\frac{S-\frac{N}{K}}{K-1}\right)d, \quad H(A^{t+1}_{i,k}|Z^{t+1}_k)=0,
 \end{align}
 for $k,i\in\{1,\ldots,K\}$, and $i\neq k$. 
 
We next define the worst-case communication as follows: 
\begin{definition}[\textbf{Worst-Case Communication}] For any achievable scheme characterized by the functions $(\phi,\psi,\mu)$, the worst-case communication overhead over all possible consecutive data shuffles $(\pi_{t}, \pi_{t+1})$ is defined  as
\begin{equation}
R_{\textsf{worst-case}}^{(\phi,\psi,\mu)}(K,S)=\underset{(\pi_{t},\pi_{t+1})}{\max}\; R_{(\pi_{t}, \pi_{t+1})}^{(\phi,\psi,\mu)}(K,S).
\end{equation}
\end{definition}

Our goal in this work is to characterize the optimal worst-case communication $R_{\textsf{worst-case}}^*(K,S)$ defined as
\begin{equation}
R_{\textsf{worst-case}}^*(K,S)=\underset{(\phi,\psi,\mu)}{\min} \;R_{\textsf{worst-case}}^{(\phi,\psi,\mu)}(K,S).
\end{equation}
We next present a claim which shows that the optimal communication $R^*$ (for any shuffle including the worst-case) is a convex function of the storage $S$:
\begin{claim}
\label{cl:1}
$R^*$ is a convex function of $S$, where $S$ is the available storage at each worker.
\end{claim}
\vspace{-10pt}

\begin{proof}
Claim~\ref{cl:1} follows from a simple memory sharing argument which shows that for any two available storage values $S_1$ and $S_2$, if $(S_1,R^*(K,S_1))$, and $(S_2,R^*(K,S_2))$ are achievable optimal schemes, then for any storage $\bar{S}=\alpha S_1 +(1-\alpha) S_2$, $0\leq\alpha\leq1$, there is a scheme which achieves a communication overhead of $\bar{R}=\alpha R^*(K,S_1) +(1-\alpha) R^*(K,S_2)$. 

This is done as follows: First, we divide the data-set $A$ across $d$ dimensions into 2 batches namely; $A^{(\alpha)}$, and $A^{(1-\alpha)}$ of dimensions $\alpha d$, and $(1-\alpha) d$, for each point respectively. Then, we divide the storage for every worker $w_k$ into 2 parts namely; $Z_k^{(\alpha)}$, and $Z_k^{(1-\alpha)}$ of size $S_1\alpha d$, and $S_2(1-\alpha)d$, respectively. The former batch $A^{(\alpha)}$ will be shuffled among the former part of the storage $Z_k^{(\alpha)}$ to achieve the point $(S_1,R^*(K,S_1))$, while the latter batch $A^{(1-\alpha)}$ will be shuffled among the latter part of the storage $Z_k^{(1-\alpha)}$ to achieve the point $(S_2,R^*(K,S_2))$. Therefore, the total achievable load is given by
\begin{equation}
\label{eq:memory-sharing}
H(X)=R^*(K,S_1) \alpha d+ R^*(K,S_2) (1-\alpha) d= \bar{R}d.
\end{equation}
We next note that the optimal communication rate $R^*(K,\bar{S})$ is upper bounded by $\bar{R}(K,\bar{S})$, the rate of the memory sharing scheme, which completes the proof.
\end{proof}


\vspace{-5pt}
\section{Main Results}
\label{sec:Results}
\begin{theorem}
\label{th:1}
For the distributed shuffling problem with $K=2$ workers of storage $Sd$ bits each, and a data-set of size $Nd$ bits, the optimal communication versus storage tradeoff is given by
\begin{equation}
\label{eq:thm1}
 R^*_{\textsf{worst-case}}(2,S) = N-S, \quad \frac{N}{2}\leq S\leq N .
\end{equation}
\end{theorem}

\begin{figure}[t]
  \begin{center}
  \includegraphics[width=0.65\columnwidth]{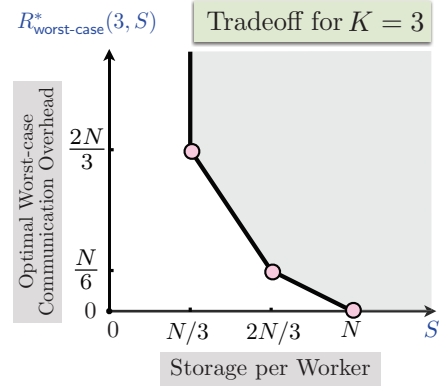}
\caption{The worst-case communication overhead is depicted versus different values of storage for $K=3$ workers. The shaded grey region is the information theoretically optimal tradeoff.}
\label{fig:theorem}
  \end{center}
\vspace{-15pt}
\end{figure}

\begin{theorem}
\label{th:2}
For a distributed shuffling problem system with $K=3$ workers, the optimal communication versus storage tradeoff is given by
\begin{equation}
 R^*_{\textsf{worst-case}}(3,S) =\left\{ \begin{array}{c c}
 \frac{7N}{6}-\frac{3S}{2}, &\frac{N}{3}\leq S\leq \frac{2N}{3}\\
 \frac{N}{2}-\frac{S}{2},  &\frac{2N}{3}< S\leq N
\end{array}.
\right.
\end{equation}
\end{theorem}
One interesting implication of Theorem \ref{th:2} for $K=3$ workers is that the corner point $(\frac{2N}{3},\frac{N}{6})$ as in Fig.~\ref{fig:theorem} is better than memory sharing between the two points $(\frac{N}{3},\frac{2N}{3})$, and $(N,0)$, which falls on the line connecting the two points, i.e., memory sharing here is not optimal, and coding can be leveraged to reduce the communication overhead.
%


%

\section{Proof of Theorem \ref{th:1} \small{($K=2$ workers)}}\label{sec:Theorem1}
\subsection{Achievability for $K=2$ workers}
\label{subsec:Achieve-2w}
We start with the achievablity of the corner points $S=N$ and $S=\frac{N}{2}$. The point $S=N$ is trivial and represents the case when the workers can store the whole data-set. In this case, no communication is necessary, i.e.,  $(S, R^{*})= (N,0)$ is achievable. 

The point corresponding to $S=\frac{N}{2}$ is not trivial, where each worker can only store half of the data-set. Let us assume the data batches at time $t$ for the workers $w_1$, and $w_2$ are $A^{t}_1$, and $A^{t}_2$, respectively. These batches should be stored at the corresponding workers, which are just enough to store them, i.e., $Z^{t}_1=A^{t}_1$, and $Z^{t}_2=A^{t}_2$. Recall that these batches are equally sized of $\frac{N}{2}d$ bits. For the next iteration $t+1$, the data is randomly shuffled at the master node such that the new batches are $A^{t+1}_1$, and $A^{t+1}_2$, also of the same equal size $\frac{N}{2}d$. In the delivery phase, the master node sends
\begin{align}
\label{eq:2workers}
X_{(\pi_t,\pi_{t+1})}= A^{t}_1\oplus A^{t}_2.
\end{align}
worker $w_1$ uses $X_{(\pi_t,\pi_{t+1})}$ and $Z^{t}_1=A^{t}_1$ to decode $A^{t}_2$, and get access to the whole data-set $A=\{A^{t}_1,A^{t}_2\}=\{A^{t+1}_1,A^{t+1}_2\}$. The storage update is only storing the desired batch $A^{t+1}_1$. The same procedure applies for $w_2$. Note that this choice of the transmitted function in (\ref{eq:2workers}) works for any possible shuffling, which gives a constant communication load $H(X_{(\pi_t,\pi_{t+1})})=\frac{N}{2}d$, and hence the corner point $(\frac{N}{2},\frac{N}{2})$ is achievable.

 With the achievability of the corner points, any point in between for any real number $S$ can be simply achieved by memory sharing (see Claim \ref{cl:1}). 
Therefore, the upper bound for the optimal worst-case communication rate is given by
\begin{align}
\label{eq:2worker-upper}
 R_{\textsf{worst-case}}^*(2,S) \leq N-S, \quad \frac{N}{2}\leq S\leq N .
\end{align}

\begin{remark}[\textbf{From worst-case to any shuffle}]
\normalfont
If there is an overlap between $A^{t}_k$ and $A^{t+1}_k$ for $k\in\{1,2\}$, then for $S=\frac{N}{2}$, the  communication cost of the above scheme can be further improved by sending
\begin{equation}
\label{eq:2workers-modified}
X_{(\pi_t,\pi_{t+1})}^{\textsf{any-shuffle}}= (A^{t}_1\setminus A^{t+1}_1) \oplus (A^{t}_2\setminus A^{t+1}_2),
\end{equation}
where $A^{t}_k\setminus A^{t+1}_k$ represents the part of the old batch $A^{t}_k$ at worker $w_k$, which is not needed any more in the new batch $A^{t+1}_k$. 
For an overlap $|A^{t}_k \cap A^{t+1}_k|=b$ data points, where $b$ is an integer number $b \in \{0,\ldots,\frac{N}{2}\}$, then the we achieve the corner point $(\frac{N}{2},\frac{N}{2}-b)$, with the worst-case rate when $b=0$. \end{remark}

\subsection{Converse for $K=2$ workers}
In this section, we present an information theoretic lower bound for the worst-case communication rate which matches the above scheme for $K=2$ workers. 
\begin{remark}[\textbf{Basic idea for the converse}]\normalfont
\label{re:basic-idead-converse}
Since we do not know a priori the shuffle that gives the worst-case communication, we assume first a shuffle $(\pi_t,\pi_{t+1})$ with the optimal rate $R_{(\pi_t,\pi_{t+1})}^*$, and then we lower bound  $R_{(\pi_t,\pi_{t+1})}^*$. Since the optimal worst-case rate is larger than the rate for any shuffle, i.e., $R_{\textsf{worst-case}}^*\geq R_{(\pi_t,\pi_{t+1})}^*$,  the lower bound found over $R_{(\pi_t,\pi_{t+1})}^*$ serves also as a lower bound for the worst-case communication $R_{\textsf{worst-case}}^*$. The novel part in our proof is choosing the right shuffle which leads to the optimal lower bound (also see Section~\ref{sec:converse-Th2} for the application of this idea to the converse proof for Theorem $2$).
\end{remark}

Now let us assume the data is shuffled such that $A^{t+1}_1= A^{t}_{2}$, and $A^{t+1}_2= A^{t}_{1}$, then from decodability constraint in (\ref{eq:decoding-const}):
\begin{equation}
\label{eq:decoding-const2}
H(A^{t+1}_{1}|Z^{t}_1, X_{(\pi_t,\pi_{t+1})})=H(A^{t}_{2}|Z^{t}_1, X_{(\pi_t,\pi_{t+1})})=0.
\end{equation}
Hence, we can find that $H(A|Z^{t}_1, X_{(\pi_t,\pi_{t+1})}) =0$ as follows
\begin{align}
\label{eq:decoding-const3}
&H(A|Z^{t}_1, X_{(\pi_t,\pi_{t+1})})\overset{(a)}{=}H(A^{t}_1,A^{t}_{2}|Z^{t}_1, X_{(\pi_t,\pi_{t+1})}) \nonumber\\ 
&\overset{(b)}{\leq} H(A^{t}_1|Z^{t}_1, X_{(\pi_t,\pi_{t+1})})+H(A^{t}_{2}|Z^{t}_1, X_{(\pi_t,\pi_{t+1})}) \nonumber\\
&\overset{(c)}{\leq}  H(A^{t}_1|Z^{t}_1)+H(A^{t+1}_{1}|Z^{t}_1, X_{(\pi_t,\pi_{t+1})})\overset{(d)}{=}0,
\end{align}
where $(a)$ follows from (\ref{eq:data-batches}), $(b)$ follows from the fact that $H(A,B|C) \leq H(A|C)+H(B|C)$, $(c)$ is because conditioning reduces entropy, and $(d)$ is from  (\ref{eq:cache-min-desired}) and (\ref{eq:decoding-const2}). 
We next prove the lower bound as follows
\begin{align}
Nd&\overset{(a)}{=} H(A)\overset{(b)}{=} H(A|Z^{t}_1,X_{(\pi_t,\pi_{t+1})})+I(A;Z^{t}_1,X_{(\pi_t,\pi_{t+1})} )\nonumber\\
&\overset{(c)}{=} H(Z^{t}_1,X_{(\pi_t,\pi_{t+1})}) -H(Z^{t}_1,X_{(\pi_t,\pi_{t+1})}|A)\nonumber\\
&\overset{(d)}{\leq} H(Z^{t}_1)+H(X_{(\pi_t,\pi_{t+1})}) \overset{(e)}{\leq} Sd +  R^*_{(\pi_t,\pi_{t+1})}d,
\end{align}
where $(a)$ follows from (\ref{eq:data-set}), $(b)$, and $(c)$ follow from the fact that $I(A;B)= H(A)-H(A|B)=H(B)-H(B|A)$ as well as (\ref{eq:decoding-const3}), $(d)$ is due to the fact that $H(A,B)\leq H(A)+H(B)$ and the fact that the $Z^{t}_1$ and $X_{(\pi_t,\pi_{t+1})}$ are functions of the whole data-set $A$, i.e., $H(Z^{t}_1,X_{(\pi_t,\pi_{t+1})}|A)=0$, and $(e)$ follows from (\ref{eq:cache-storage}) and (\ref{eq:transmit-load}). Hence, from Remark~\ref{re:basic-idead-converse}, the lower bound for the worst-case rate is characterized as
\begin{equation}
\label{eq:2worker-lower}
 R_{\textsf{worst-case}}^*(2,S) \geq N-S, \quad\frac{N}{2}\leq S\leq N.
\end{equation}
Hence, the proof of Theorem \ref{th:1} is complete from (\ref{eq:2worker-upper}) and (\ref{eq:2worker-lower}).

\vspace{-6pt}

\section{Proof of Theorem \ref{th:2} \small{($K=3$ workers)}}\label{sec:Theorem2}
\subsection{Achievability for $K=3$ workers}
\label{subsec:Achieve-3w}
From Fig.~\ref{fig:theorem}, the achievability involves three corner points: $S=\frac{N}{3}$, $S=\frac{2N}{3}$, and $S=N$. Achieving the point $(N,0)$ is trivial. The scheme for the point $S=\frac{N}{3}$ is similar to the point $S=\frac{N}{2}$ in the $K=2$ worker case, where each worker can only store the desired data batches. Similar to (\ref{eq:2workers}) the transmission at time $t+1$ is 
$X_{(\pi_t,\pi_{t+1})}= \{A^{t}_1\oplus A^{t}_2, A^{t}_2\oplus A^{t}_3\}$,
which is sufficient for each worker to access all the data-set and store what it needs, achieving the corner point $(\frac{N}{3},\frac{2N}{3})$. 

In this section, we focus on the achievability of the corner point $\left( \frac{2N}{3}, \frac{N}{6}\right)$, which is perhaps the most interesting aspect of this result. For each worker $w_k$ at time $t$, $k\in \{1,2,3\}$, half of the storage ($N/3$ points) is used to store the desired batch $A^{t}_k$, while the remaining half (excess storage of $N/3$ points) is used opportunistically in order to minimize the communication overhead by storing some parts of the remaining batches given by the sub-batches $A^{t}_{i,k},\;  i\in\{1,2,3\}\setminus k$. 

%

\begin{figure*}
\begin{center}
\includegraphics[width=0.8\textwidth]{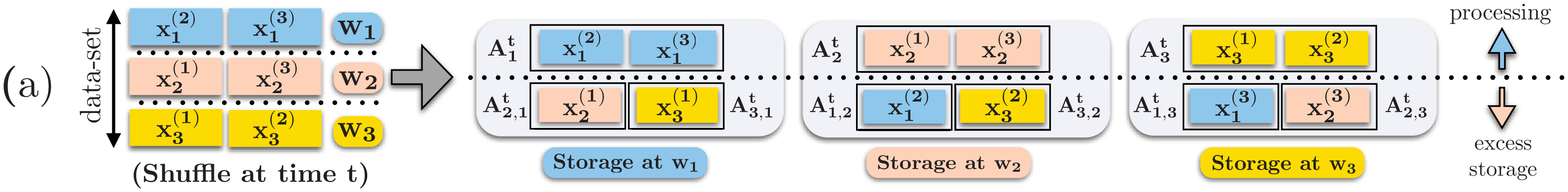}
\includegraphics[width=0.8\textwidth]{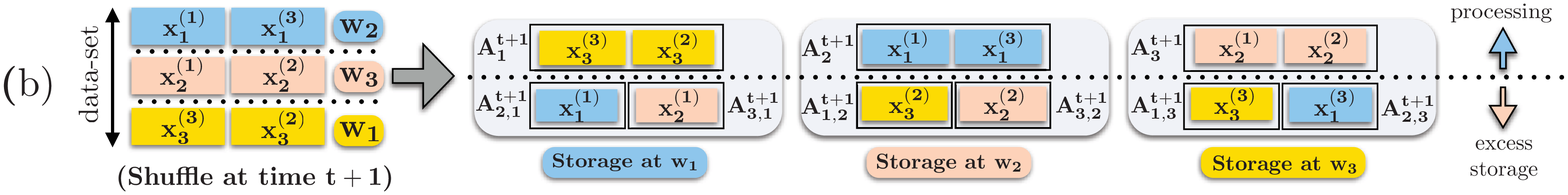}
\vspace{-5pt}
\caption{ Above the dotted line is the data under processing, and below  is the excess storage used to reduce the communication overhead. (a) The storage placement for the shuffle at time $t$ of a data-set with $N=3$ data points, for $K=3$ workers, and storage $S=\frac{2N}{3}=2$ points. Each data points is divided into two equal sub-divisions to maintain the structurally invariant placement, such that each worker obtains the data point under processing as well as half of each of the two remaining points. (b) The storage update at time $t+1$. We can notice that this process maintains the structural properties. In order to update the storage at time $t+1$, we need the delivery of the coded symbol $\{x_3^{(2)}\oplus x_1^{(3)} \oplus x_2^{(1)}\}$. \label{fig:example-3worker}}
\label{fig:example-3worker}
\vspace{-17pt}
  \end{center}
\end{figure*}

These parts will be formed as follows: if we consider the data batch $A^{t}_k$ assigned for $w_k$ of $\frac{N}{3}$ points, each point $x\in A^{t}_k$ is divided across $d$ dimensions into two equal subdivisions labeled as $x^{(i)},\;i\in\{1,2,3\}\setminus k$. Then, each one of these subdivisions $x^{(i)}$ is placed in the corresponding sub-batch $A^{t}_{k,i}$ (the part of $x\in A^{t}_k$ stored in the excess storage of $w_i$). 
For instance, $A^{t}_1$ (of size $\frac{N}{3}$) stored in the processing half of the storage of worker $w_1$ is divided into two equal non-overlapping parts $A^{t}_1 = \{A^{t}_{1,2}, A^{t}_{1,3}\}$ (of size $\frac{N}{6}$ each). Worker $w_2$ will use half of its excess storage to store $A^{t}_{1,2}$, while worker $w_3$ will use half of its excess storage to store $A^{t}_{1,3}$.

The storage update procedure we present here maintains the above structural property of the stored data over time. The consequence of such structural invariance is that for any data point is required to be at a worker, at least half of this point is guaranteed to be already present at the worker, which decreases the communication overhead of the shuffling process.

\noindent \underline{\textbf{Data Delivery:}} With this placement strategy, for the subsequent shuffles $(\pi_t,\pi_{t+1})$, the transmitted function is given as
 \begin{equation}
\label{eq:3workers-point3}
X_{(\pi_t,\pi_{t+1})}= (A^{t+1}_1\setminus Z^{t}_1) \oplus (A^{t+1}_2\setminus Z^{t}_2) \oplus (A^{t+1}_3\setminus Z^{t}_3).
\end{equation}

We claim that the above transmission is sufficient for all the three workers to obtain the required new points for any shuffle. Without loss of generality, let us consider worker $w_1$. According to (\ref{eq:3workers-point3}), for $w_1$ to obtain the needed points not available in its storage, $A^{t+1}_1\setminus Z^{t}_1$, it must have $A^{t+1}_2\setminus Z^{t}_2$ and $A^{t+1}_3\setminus Z^{t}_3$ in its storage $Z_{1}^{t}$. This is indeed the case and can be proved according to the following argument: 


In the following, we prove that $A^{t+1}_2\setminus Z^{t}_2 \in Z_{1}^{t}$, and using the same argument we can show that $A^{t+1}_3\setminus Z^{t}_3 \in Z_{1}^{t}$.  Consider a data point $x \in A_2^{t+1}$, which is newly assigned to worker $w_2$ at time $t+1$ and is not fully present in its storage $Z_{2}^{(t)}$, i.e., $x\not\in A_2^{t}$. Therefore, there are two possibilities: a) $x\in A_1^{t}$ was being processed by worker $w_1$ at time $t$, which directly implies that $x$ is already available at $w_1$; or
b) $x\in A_3^{t}$ was being processed by worker $w_3$ at time $t$, which implies that the sub-divisions of $x$ are $\{x^{(1)}, x^{(2)}\}$ and the needed part by $w_2$, $x^{(1)}=x\cap (A^{t+1}_2\setminus Z^{t}_2)$, is available at $w_1$ by definition (since $x^{(2)}$ is the subdivision that is already present at $w_2$).


According to (\ref{eq:3workers-point3}), the worst-case scenario for this scheme happens if there is no overlap between $A^{t}_k$ and $A^{t+1}_k$ (completely new assignments). However, half of each data point $x\in A^{t+1}_k$ is already stored in the excess storage at $w_k$, labeled $x^{(k)}$, then the worst-case rate of $A^{t+1}_k\setminus Z^{t}_k$ and eventually $X_{(\pi_t,\pi_{t+1})}$ is $\frac{N}{6}$, achieving the corner point $(\frac{2N}{3},\frac{N}{6})$.

\noindent \underline{\textbf{Storage Update:}} Now, we present a deterministic storage update strategy, which maintains the structural properties of the storage at time $t$.
Without loss of generality, let us analyze a data point $x=\{x^{(2)},x^{(3)}\}\in A^{t+1}_1$ . The update procedure is done according to the following three cases at time $t$

\noindent $\bullet$ \textbf{Case 1:} ${ x=\{x^{(2)},x^{(3)}\}\in A^{t}_1}$. In this case, since the point $x$ was already being processed at worker $w_1$ at time $t$, hence no storage update is necessary at time $t+1$ for this point across workers.

\noindent $\bullet$  \textbf{Case 2:} $x=\{x^{(1)},x^{(3)}\}\in A^{t}_2$,  $x^{(1)}\in A^{t}_{2,1}$, $x^{(3)}\in A^{t}_{2,3}$.
 After receiving $x^{(3)}$ from the delivery phase, worker $w_1$ stores the full-point $x$ in $A^{t+1}_1$. 
For worker $w_3$, $x^{(3)}$ leaves $A^{t}_{2,3}$ and enters $A^{t+1}_{1,3}$, and we can notice that $x^{(3)}$ remains within the excess storage of $Z_3^{t+1}$. Worker $w_2$ removes the data point $x$ from the processing batch $A^{t+1}_2$, and stores $x^{(1)}$ in $A_{1,2}^{t+1}$ after relabelling it as $x^{(2)}$ (now stored at the excess storage of $w_2$), i.e., it simply moves one half of $x$ in its excess storage. 

\noindent $\bullet$ \textbf{Case 3:} $x=\{x^{(1)},x^{(2)}\}\in A^{t}_3$,  $x^{(1)}\in A^{t}_{3,1}$, $x^{(2)}\in A^{t}_{3,2}$.
This case is similar to case 2, where $w_1$ stores $x$ in $A^{t+1}_1$ and removes $x^{(1)}$ from $A^{t+1}_{3,1}$, worker $w_2$ moves $x^{(2)}$ into $A^{t+1}_{1,2}$ instead of $A_{3,2}^{t}$, and worker $w_3$ removes $x$ from $A^{t+1}_{3}$ and stores $x^{(1)}$ (now labeled as $x^{(3)}$) in $A^{t+1}_{1,3}$.

 \begin{example}[$\mathbf{N=3, \: K=3,\: S=\frac{2N}{3}}=2$]
 \normalfont
Let us now take a representative example depicted in Fig.~\ref{fig:example-3worker} to illustrate our proposed data-delivery and storage-update phases for the corner point $(\frac{2N}{3},\frac{N}{6})$.
Consider a system with $K=3$ workers, and $N=3$ data points, $\{x_1,x_2,x_3\}$.
We assume that storage per worker is $S=\frac{2N}{K}=2$ points, i.e., each worker can store one extra data point in addition to the one under processing.
We first clarify the color code used in this example to indicate the data point assigned to a certain worker labeled with these colors:  
blue for $x_1$, red for $x_2$, and yellow for $x_3$.

 At time $t$, consider the dataset is shuffled such that $A_1^{t}= x_1$, $A_2^{t}= x_2$, and $A_3^{t}= x_3$. The corresponding storage placement at time $t$ in Fig.~ \ref{fig:example-3worker}(a) is as follows: After using half the storage  to store the desired data point (which can be depicted in this example as the desired color), each data point is divided equally among the unintended workers (depicted in this example as the unintended colors). 
 For example, if we take the batch $A_1^{t}= x_1=\{x^{(2)}_1,x^{(3)}_1\}$, worker $w_2$ stores $A_{1,2}^{t}=x_1^{(2)}$, and worker $w_3$ stores $A_{1,3}^{t}=x_1^{(3)}$.
 
At time $t+1$ in Fig.~\ref{fig:example-3worker}(b), the data is randomly shuffled again such that the new batches are: $A_1^{t+1}= x_3$, $A_2^{t+1}= x_1$, and $A_3^{t+1}= x_2$. We take this particular shuffle since it represents one of the possible worst cases, where every worker is assigned a completely different batch. According to the previous storage content shown in Fig.~\ref{fig:example-3worker}(a), worker $w_1$ already has $x_3^{(1)}$ but still needs $x_3^{(2)}$, which is stored at $w_2$ and $w_3$. Similarly, worker $w_2$ needs $x_1^{(3)}$ which is stored at $w_1$ and $w_3$, and worker $w_3$ needs $x_2^{(1)}$ which is stored at $w_1$ and $w_2$. Following the data delivery as in (\ref{eq:3workers-point3}), the master node transmits: 
\begin{equation}
X_{(\pi_t,\pi_{t+1})}=x_3^{(2)}\oplus x_1^{(3)} \oplus x_2^{(1)}.
\end{equation}
Each worker has two out of these three subdivisions, therefore it can decode the remaining needed one.
 The rate of this transmission is  $R=\frac{1}{2}$, which is $\frac{N}{6}$ where $N=3$.


For the storage update in Fig.~\ref{fig:example-3worker}(b), we only discuss the changes in $A^{t+1}_1$ and the corresponding $A^{t+1}_{1,2}$, and $A^{t+1}_{1,3}$. The storage update of the remaining parts can be done in a similar manner. 
From the delivery phase and the previous storage, worker $w_1$ gets $x_3$ (labeled yellow) and stores it in the processing half $A^{t+1}_1$ (above the dotted line). Worker $w_2$ already has a part of $A^{t+1}_1$, $x^{(2)}_3$ (labeled yellow), which was previously stored as $A^{t}_{3,2}$. Therefore, it remains in the excess storage (below the dotted line) as $A^{t+1}_{1,2}$. Worker $w_3$ already has $x_3$ previously labeled as $A^{t}_3$, so it keeps in its excess storage the part that is not stored in $A^{t+1}_{1,2}$, i.e., $x_3^{(1)}$,  to be stored in $A^{t+1}_{1,3}$, after relabelling it to $x_3^{(3)}$ (now stored in $w_3$). 
  \end{example} 
Using the achievability of the three corner points for $K=3$, and Claim~\ref{cl:1}, we get an upper bound on $R^*_{\textsf{worst-case}} (3,S)$ as 
\begin{equation}
 R^*_{\textsf{worst-case}} (3,S) \leq \left\{ \begin{array}{c c}
 \frac{7N}{6}-\frac{3S}{2}, &\frac{N}{3}\leq S\leq \frac{2N}{3}\\
 \frac{N}{2}-\frac{S}{2}, &\frac{2N}{3}< S\leq N
\end{array}.
\right.
\label{eq:upper-3w}
\end{equation}

\subsection{Converse for $K=3$ workers}\label{sec:converse-Th2}
We now present the information theoretic lower bounds for the three-worker case, which matches the above scheme. Following Remark~\ref{re:basic-idead-converse}, we first assume subsequent data shuffles at times $\{t,t+1,t+2\}$ such that $A^{t+1}_1= A^{t}_{2}$, and $A^{t+2}_1= A^{t}_{3}$, then from the decodability constraint in (\ref{eq:decoding-const}), we have
\begin{equation}
\label{eq:decoding-const3w}
H(A^{t}_{2}|Z^{t}_1, X_{(\pi_t,\pi_{t+1})})=H(A^{t}_{3}|Z^{t+1}_1, X_{(\pi_{t+1},\pi_{t+2})})=0.
\end{equation}
Hence, in a similar proof to (\ref{eq:decoding-const3}) we get
\begin{align}
\label{eq:decoding-const3w-0}
H(&A|Z^{t}_1, X_{(\pi_t,\pi_{t+1})}, X_{(\pi_{t+1},\pi_{t+2})})\nonumber\\ 
&\overset{(a)}{=}H(A^t_1,A^t_2,A^t_3|Z^{t}_1,Z^{t+1}_1, X_{(\pi_t,\pi_{t+1})}, X_{(\pi_{t+1},\pi_{t+2})}) \nonumber\\ 
&\overset{(b)}{\leq} H(A^{t}_1|Z^{t}_1)+H(A^{t}_{2}|Z^{t}_1, X_{(\pi_t,\pi_{t+1})})+\nonumber\\
&\qquad H(A^{t}_{3}|Z^{t}_1,Z^{t+1}_1, X_{(\pi_t,\pi_{t+1})},X_{(\pi_{t+1},\pi_{t+2})}) \nonumber\\
&\overset{(c)}{=} H(A^{t}_{3}|Z^{t}_1,Z^{t+1}_1, X_{(\pi_t,\pi_{t+1})},X_{(\pi_{t+1},\pi_{t+2})}) \overset{(d)}{=} 0,
\end{align}
 where $(a)$ follows from (\ref{eq:data-batches}) and  the storage update constraint in  (\ref{eq:cache-update}), $(b)$ follows from the fact that $H(A,B,C|D) \leq H(A|D)+H(B|D)+H(C|D)$, and also the fact that conditioning reduces entropy, $(c)$ from (\ref{eq:cache-min-desired}) and (\ref{eq:decoding-const3w}), and $(d)$ from (\ref{eq:decoding-const3w}). Now, using (\ref{eq:decoding-const3w-0})  we can find the upper bound as follows
 \begin{align}
Nd&\overset{(a)}{=} H(A)\overset{(b)}{=} I(A;Z^{t}_1, X_{(\pi_t,\pi_{t+1})},X_{(\pi_{t+1},\pi_{t+2})} )\nonumber\\
&\overset{(c)}{=} H(Z^{t}_1, X_{(\pi_t,\pi_{t+1})},X_{(\pi_{t+1},\pi_{t+2})} ) \nonumber\\
&\qquad-H(Z^{t}_1, X_{(\pi_t,\pi_{t+1})},X_{(\pi_{t+1},\pi_{t+2})}|A)\nonumber\\
&\overset{(d)}{\leq} H(Z^{t}_1)+H(X_{(\pi_t,\pi_{t+1})} )+H(X_{(\pi_t,\pi_{t+1})} )\nonumber\\ 
&\overset{(e)}{\leq} Sd +  R^*_{(\pi_t,\pi_{t+1})} d+R^*_{(\pi_{t+1},\pi_{t+2})}d,
\end{align}
where $(a)$ follows from (\ref{eq:data-set}), $(b)$, and $(c)$ follow from (\ref{eq:decoding-const3w-0}), and due to the fact that $I(A;B)= H(A)-H(A|B)=H(B)-H(B|A)$, $(d)$ is due to the fact that $H(A,B,C)\leq H(A)+H(B)+H(C)$ and the fact that $Z^{t}_1$, $X_{(\pi_t,\pi_{t+1})} $, and $X_{(\pi_{t+1},\pi_{t+2})} $ are all functions of the whole data-set $A$, and $(e)$ follows from (\ref{eq:cache-storage}) and (\ref{eq:transmit-load}).  Hence, following Remark~\ref{re:basic-idead-converse}, we get a lower bound for the worst-case rate characterized as
 \begin{equation}
\label{eq:lower-3w}
 R^*_{\textsf{worst-case}}(3,S) \geq \frac{N-S}{2}\quad, \frac{N}{3}\leq S\leq N.
\end{equation}
The lower bound in (\ref{eq:lower-3w}) matches the upper bound obtained in (\ref{eq:upper-3w}) for the range $\frac{2N}{3}\leq S\leq N$.

We next present another lower bound on $R^*(3,S)$ which proves the optimality of our scheme for the range $\frac{N}{3}\leq S\leq \frac{2N}{3}$. 
To this end, we now assume a data shuffle such that $A_1^{t+1}=A_3^{t}$.
Similar to (\ref{eq:decoding-const3}) and (\ref{eq:decoding-const3w-0}), we have
\begin{align}
\label{eq:decoding-const3w-3}
&H(A|Z^{t}_1,Z^{t}_2, X_{(\pi_{t},\pi_{t+1})})\overset{(a)}{=}H(A^{t}_1,A^{t}_2,A^{t}_3|Z^{t}_1,Z^{t}_2, X_{(\pi_{t},\pi_{t+1})}) \nonumber\\
 &\overset{(b)}{\leq} H(A^{t}_1|Z^{t}_1)+H(A^{t}_{2}|Z^{t}_2)+H(A^{t}_{3}|Z^{t}_1, X_{(\pi_{t},\pi_{t+1})}) \overset{(c)}{=} 0,
\end{align}
where $(a)$ follows from (\ref{eq:data-batches}), $(b)$  from the facts that $H(A,B,C|D) \leq H(A|D)+H(B|D)+H(C|D)$, and $(c)$ using (\ref{eq:cache-min-desired}), and  (\ref{eq:decoding-const}). We now proceed to obtain the second lower bound on $R^*_{\textsf{worst-case}}$ as follows
 \begin{align*}
 \label{eq:lower-3w-2}
&Nd\overset{(a)}{=} H(A) \overset{(b)}{=} I(A;Z^{t}_1,Z^{t}_2, X_{(\pi_{t},\pi_{t+1})})\nonumber\\
&\overset{(c)}{=} H(Z^{t}_1,Z^{t}_2, X_{(\pi_{t},\pi_{t+1})}) -H(Z^{t}_1,Z^{t}_2, X_{(\pi_{t},\pi_{t+1})} |A)\nonumber\\
&\overset{(d)}{=} H(Z^{t}_1, X_{(\pi_{t},\pi_{t+1})})+H(Z^{t}_2|Z^{t}_1, X_{(\pi_{t},\pi_{t+1})})\nonumber \\
&\overset{(e)}{=} H(Z^{t}_1, X_{(\pi_{t},\pi_{t+1})})+ H(Z^{t}_2|Z^{t}_1, X_{(\pi_{t},\pi_{t+1})},A_1^{t},A_3^{t}, A_{2,1}^{t})\nonumber \\
&\overset{(f)}{\leq} H(Z^{t}_1, X_{(\pi_{t},\pi_{t+1})})+ H(Z^{t}_2|A_1^{t},A_3^{t}, A_{2,1}^{t})\nonumber \\
&\overset{(g)}{=} H(Z^{t}_1, X_{(\pi_{t},\pi_{t+1})})+ H(A_{2,3}^{t})\nonumber \\
\end{align*}
\begin{align}
&\overset{(h)}{\leq} Sd +  R^*_{(\pi_t,\pi_{t+1})}d+ \frac{S-\frac{N}{3}}{2}d,
\end{align}
where $(a)$ follows from (\ref{eq:data-set}),
 $(b)$, and $(c)$ from (\ref{eq:decoding-const3w-3}), and due to the fact that $I(A;B)= H(A)-H(A|B)=H(B)-H(B|A)$, $(d)$ from the chain rule of entropy and the fact that $Z^{t}_1$, $Z^{t}_2$, and $X_{(\pi_t,\pi_{t+1})}$ are all functions of the whole data-set $A$, $(e)$ from  (\ref{eq:decoding-const}) where $A_1^{t+1}=A_3^{t}$ must be decoded from $Z_{1}^{t}$, and $X_{(\pi_t,\pi_{t+1})}$, and because $A_1^{t}$ and $A^{t}_{2,1}$ are stored within $Z^{t}_1$, $(f)$ because conditioning reduces entropy, $(g)$ because after obtaining $A_1^{t}$, $A_3^{t}$, and $A_{2,1}^{t}$, the only remaining part in $Z_2^{t}$ is $A_{2,3}^{t}$, and finally $(h)$ follows from (\ref{eq:cache-storage}), (\ref{eq:transmit-load}), and (\ref{eq:excess-storage}).
From Remark~\ref{re:basic-idead-converse}, and by rearranging (\ref{eq:lower-3w-2}), we get the following bound
 \begin{equation}
\label{eq:lower-3w-3}
 R^*_{\textsf{worst-case}}(3,S) \geq \frac{7N}{6}-\frac{3S}{2}, \quad\frac{N}{3}\leq S\leq N.
\end{equation}
Therefore, from (\ref{eq:lower-3w-3}), and (\ref{eq:lower-3w}), we get the following lower bound on $R^*_{\textsf{worst-case}}(3,S)$:
\begin{equation}
 R^*_{\textsf{worst-case}}(3,S)  \geq \left\{ \begin{array}{c c}
 \frac{7N}{6}-\frac{3S}{2},&\frac{N}{3}\leq S\leq \frac{2N}{3}\\
 \frac{N}{2}-\frac{S}{2}, &\frac{2N}{3}< S\leq N
\end{array}
\right.
\label{eq:lower-3w-4}
\end{equation}
Finally, the proof of Theorem \ref{th:2} follows from (\ref{eq:upper-3w}) and (\ref{eq:lower-3w-4}).

\section{Conclusions}
In this paper, we presented information theoretic formulation of the data shuffling problem, where we studied the tradeoff between the worst-case communication overhead and the storage available at the worker nodes.
  We completely characterized the optimal worst-case communication for $K=2$, and $K=3$ workers with any storage capacity, where we leveraged excess storage and coding to minimize the communication overhead in subsequent data shuffling iteration.
  A systematic storage update and delivery scheme was presented, which preserves the structural properties of the storage across workers.
  Generalizing these results for any number of workers $(K>3)$ is part of our ongoing work. 

\label{sec:conclusion}
\vspace{-6pt}

%

%
\bibliographystyle{IEEEtran}
\nocite{*}
\bibliography{Final-GC2016.bbl}

\end{document}